\documentclass[11pt]{article}
\usepackage{moriond,epsfig}
\usepackage{amssymb}

\def\be{\begin{equation}}
\def\ee{\end{equation}}
\def\bea{\begin{eqnarray}}
\def\eea{\end{eqnarray}}

\begin{document}
\vspace*{4cm}
\title{TOP-QUARK FORWARD-BACKWARD SYMMETRY}

\author{SUSANNE WESTHOFF}

\address{Institut f\"ur Physik (THEP),
Johannes Gutenberg-Universit\"at \\
D-55099 Mainz, Germany}

\maketitle\abstracts{
Top-quark pair production at the Tevatron is discussed within the Randall-Sundrum model of warped extra dimensions. Generically, the exchange of massive Kaluza-Klein gluons has the potential to generate a large forward-backward asymmetry. In models with an anarchic flavor structure, however, their coupling to the light quarks inside the proton is strongly suppressed. The consequent suppression of the asymmetry at tree level is lifted at next-to-leading order. Still, it is not possible to increase the forward-backward asymmetry with respect to the Standard-Model prediction in this framework.}

\section{Introduction}
Since the discovery of the top quark in 1995, the experiments at the Tevatron have made remarkable achievements in measuring the properties of the heaviest fermion in the Standard Model (SM). The motivation behind this effort is evident: due to its particularly large mass, the top quark is supposed to play a key role in understanding the mechanism of electroweak symmetry breaking. Within the SM, all quarks possess the same gauge couplings, which prevents us from explaining their strong mass hierarchy by a fundamental interaction. Extensions of the SM addressing the origin of quark masses commonly imply new gauge interactions that distinguish the heavy top quark from the light quarks. We will concentrate on massive color-octet gauge bosons with strong couplings to top quarks, which occur in a wide class of models. Examples include colorons in topcolor models, axigluons from flavor non-universal chiral color, and Kaluza-Klein (KK) gluons in warped extra dimensions. They all share the feature to yield characteristic effects in top-quark observables, which serve as probes of the underlying theory. Our framework will be the Randall-Sundrum (RS) model, where the fermion mass hierarchy  can be explained by the localization of fermion fields in the bulk of a warped extra dimension. These localizations manifest themselves in flavor- and chirality-specific couplings of quarks to KK excitations of the gluon. Since the coupling to top quarks in this setup is typically strong, the exchange of a massive KK gluon is expected to affect top-quark pair production.

\begin{figure}[t]
\begin{center}
\epsfig{figure=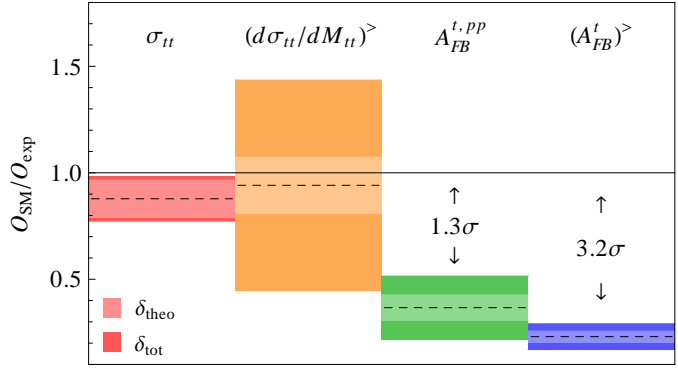,height=1.9in}
\end{center}
\caption{Top-quark pair production in the SM. The ratio $O_{\rm SM}/O_{\rm exp}$ is shown for the total cross section $\sigma_{t\bar t}$, the distribution for $M_{t\bar t}\in[0.8,1.4]\,\rm{TeV}$, the inclusive asymmetry in the lab frame $A_{\rm{FB}}^{t,p\bar p}$ and for $M_{t\bar t} > 450\,\rm{GeV}$. Central values are drawn as dashed lines.
\label{fig:SMexplab}}
\end{figure}

Let us briefly summarize the status of measurements in top-antitop quark production at the Tevatron. In proton-antiproton collisions at a center-of-mass (CM) energy of $\sqrt{s}=1.96\,\rm{TeV}$, the production of $t\bar t$ pairs is dominated by the partonic process $q\bar q\rightarrow t\bar t$. The measured total cross section and its distribution with respect to the invariant mass of the top-antitop pair, \cite{CDFnotetot,Aaltonen:2009iz}
\begin{equation}\label{eq:cs}
\sigma_{t\bar t} = (7.50 \pm 0.48)\,{\rm pb}\,, \qquad  (d\sigma_{t\bar t}/d M_{t\bar t})^{M_{t\bar t}\in [0.8,1.4]\,\rm{TeV}} = (0.068\pm 0.034)\,\frac{\rm fb}{\rm GeV}\,,
\end{equation}
are in good agreement with their SM predictions. In contrast, the forward-backward asymmetry of the top quark is in variance with the result in Quantum Chromodynamics (QCD): the measured total asymmetry in the laboratory frame and the result at $t\bar t$ invariant mass measured at CDF, \cite{Aaltonen:2011kc}
\begin{equation}\label{eq:afbt}
(A_{\rm{FB}}^t)_{\rm{exp}}^{p\bar p} = (15.0\pm 5.0_{\rm{stat}}\pm 2.4_{\rm{syst}})\%\,, \qquad (A_{\rm{FB}}^t)_{\rm{exp}}^{M_{t\bar t}\,>\,450\,\rm{GeV}} = (47.5\pm 11.2)\%\,,
\end{equation}
exceed the SM predictions by about $1.3\,\sigma$ and $3.2\,\sigma$, respectively. \footnote{This excess has also been observed in $t\bar t$ dilepton events at CDF \cite{CDFdileptonnote} and by the D0 collaboration \cite{D0brandnew}.} The situation is displayed in Figure~\ref{fig:SMexplab}, where we show the ratio of the SM prediction with respect to the CDF measurement for the four observables discussed above. The observed pattern suggests new physics (NP) with a large positive contribution to the asymmetry, but only little impact on the cross section.

In a theory with CP-conserving couplings, the forward-backward asymmetry is equivalent to a charge asymmetry,
\begin{equation}
A_{\rm FB}^t = \frac{\sigma_a}{\sigma_s}\,,\qquad \sigma_{a(s)} = \int_{0}^{1}\cos\theta\left[\frac{d\sigma(p\bar p\rightarrow t\bar{t} X)}{d\cos\theta} -(+) \frac{d\sigma(p\bar p\rightarrow \bar{t}t X)}{d\cos\theta}\right]\,.
\end{equation}
It requires a production amplitude that is antisymmetric under the interchange of $t$ and $\bar t$ in the final state for a fixed top-quark scattering angle $\theta$. In QCD, $t\bar t$ production is charge-symmetric at tree level. The asymmetry arises at next-to-leading order (NLO) from the interference of the tree-level gluon exchange with QCD box diagrams and from the interference of final- and initial-state radiation. Up-to-date QCD calculations predict a small value of the asymmetry in the laboratory frame, $(A_{\rm FB}^{t})_{\rm SM}^{p\bar p} = (4-5.6)\%$ \cite{Ahrens:2011mw}. A large charge asymmetry is expected to be generated by the positive interference of new physics (NP) with the SM gluon exchange at tree level. Any new particle in the $s$ channel of $t\bar t$ production ought to exhibit large axial-vector couplings to both light quarks, $g_A^q$, and top quarks, $g_A^t$, fulfilling $g_A^q\,g_A^t < 0$ for a mass of $\mathcal{O}(1\,\rm{TeV})$. The product of vector couplings, $g_V^q\,g_V^t$, however, is restricted to be small in order to fit the measured charge-symmetric cross section $\sigma_{t\bar t}$ and the spectrum $d\sigma_{t\bar t}/dM_{t\bar t}$.

Within the RS model, KK gluons have axial-vector couplings to quarks, due to the different localizations of left- and right-handed quark fields in the extra dimension. We explore the possibility of a large top-quark forward-backward asymmetry from KK gluon exchange in the RS model with an anarchic flavor structure. It will turn out that within this setup axial-vector couplings of KK gluons to top quarks are large, but strongly suppressed for light quarks. The asymmetry at tree level is thus negligibly small. We will explain how the suppression is lifted at NLO, yielding the leading RS contributions to the asymmetric cross section $\sigma_a$. These effects, however, partially cancel with tree-level contributions to the symmetric cross section $\sigma_s$, which normalizes the forward-backward asymmetry. The observable $A_{\rm{FB}}^t$ thus cannot be increased with respect to its SM prediction within the flavor-anarchic RS model.

\section{Randall-Sundrum model with flavor anarchy}\label{sec:RSfa}
The Randall-Sundrum model was originally designed to explain the large hierarchy between the electroweak and the Planck scales by gravitational red-shifting in a warped extra dimension \cite{Randall:1999ee}. If the SM fermions are allowed to propagate into the bulk of the fifth dimension, the RS model also offers a geometrical description of flavor. As anticipated, the fermion mass hierarchy is explained by different localizations in the extra dimension: the wave functions of light fermions are exponentially localized in the ultraviolet (UV), while the heavy fermions reside in the infrared (IR). The effective Yukawa couplings, resulting from wave-function overlap with the IR-localized Higgs boson, therefore exhibit an exponential hierarchy.  Starting from flavor anarchy, i.e. arbitrary five-dimensional Yukawa couplings of $\mathcal{O}(1)$, one naturally obtains a strong hierarchy of fermion masses, quantified by the warp factor of the extra dimension, $\exp(L) = \Lambda_{\rm UV}/\Lambda_{\rm IR} \approx 10^{16}$. The actual localization of fermions is determined by bulk mass parameters $c_f$. For quarks, the measured masses and CKM mixings require $c_q < -1/2$ and $c_t > -1/2$, corresponding to UV and IR localization, respectively.

The pattern of localization of the quark fields has a crucial impact on their couplings to KK gluons. Since KK excitations of the SM particles are associated with a scale $M_{\rm{KK}} \gtrsim 1\,\rm{TeV}$, virtual RS effects in top-quark pair production can be described in terms of dimension-six operators in the framework of an effective theory. The low-energy Lagrangian for the exchange of a KK gluon in the $s$ channel of $q\bar q\rightarrow t\bar t$ is given by \footnote{Further contributions of KK gluons in the $t$ channel, as well as of $Z$, $\gamma$, their KK excitations, and the Higgs boson, are numerically subleading and therefore neglected.}
\begin{equation}\label{eq:leff}
\mathcal{L}_{\rm{eff}} = \sum_q\sum_{A,B=L,R}\,C_{q\bar q}^{AB}\,Q_{q\bar q}^{AB}\,,\qquad Q_{q\bar q}^{AB} = (\bar q_A\,\gamma_{\mu}T^a\,q_A)(\bar t_B\,\gamma^{\mu}T^a\,t_B)\,,
\end{equation}
where $A,B=L,R$ denote the quark chiralities and $T^a$ are the generators of color $SU(3)$. For phenomenology, it is useful to consider the combinations of Wilson coefficients that correspond to vector and axial-vector structures,
\begin{equation}\label{eq:cvca}
C_{q\bar q}^V = \sum_{A,B=L,R} C_{q\bar q}^{AB}\,,\qquad C_{q\bar q}^A = \sum_{A\neq B} (C_{q\bar q}^{AA} - C_{q\bar q}^{AB})\,.
\end{equation}
These effective couplings depend on the overlap of the quark wave functions with the IR-localized KK gluons. The size of the couplings is governed by the quark profiles, given to good approximation by \cite{Casagrande:2008hr}
\begin{equation}\label{eq:profiles}
F^2(c_t) \approx 1 + 2 c_t \,, \qquad F^2(c_q) \approx (-1
- 2 c_q) \, e^{L \hspace{0.25mm} (2 c_q + 1)} \,.
\end{equation}
For mass parameters $c_t > -1/2$ and $c_q < -1/2$, the profile function is of $\mathcal{O}(1)$ for top quarks, but exponentially suppressed for light quarks. In terms of these quark profiles, the Wilson coefficients for KK gluon exchange from Eq.~(\ref{eq:cvca}) read \cite{Casagrande:2008hr,Bauer:2009cf}
\begin{eqnarray}
C_{q\bar q}^V & \approx & -\frac{\pi\alpha_s}{M_{\rm KK}^2}\,\left[F^2(c_{t_R})+F^2(c_{t_L})\right]\,,\\
C_{q\bar q}^A & \approx & -\frac{\pi\alpha_s}{M_{\rm KK}^2}\,L\,\left[F^2(c_{q_R})-F^2(c_{q_L})\right]\left[F^2(c_{t_R})-F^2(c_{t_L})\right]\,,
\end{eqnarray}
where $\alpha_s$ is the strong coupling constant of QCD. The vector coefficient $C_{q\bar q}^V$ is dominated by the top-quark profiles. Comparing with Eq.~(\ref{eq:profiles}), one observes that it is large and positive due to the IR localization of the top quark. We define the dimensionless coefficient $\tilde{C}_{q\bar q}^{V} \equiv 1\,{\rm{TeV}}^2\,C_{q\bar q}^{V} = \mathcal{O}(1)$. The axial-vector coefficient $C_{q\bar q}^A$ is enhanced by the warp factor $L\approx 37$. It is proportional to the difference between the profiles for left- and right-handed quarks. For top quarks, this difference may be large, yielding a sizeable axial-vector coupling $g_A^t$ to KK gluons. The axial-vector coupling of light quarks, however, is doubly suppressed: firstly, left- and right-handed light quarks' profiles ought to be largely identical, in order to accommodate the measured quark masses and mixings. Secondly, and even more severely, light quarks are localized in the UV and thereby exponentially suppressed, as stated below Eq.~(\ref{eq:profiles}). The effective axial-vector coefficient is thus strongly suppressed, $\tilde{C}_{q\bar q}^{A} \equiv 1\,{\rm{TeV}}^2\,C_{q\bar q}^{A} = \mathcal{O}(10^{-3})$.

\begin{figure}[t]
\begin{center}
\raisebox{0.2cm}{\epsfig{figure=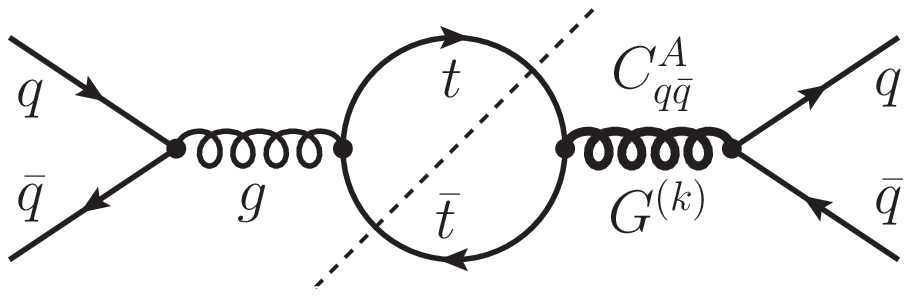,height=0.85in}}
\qquad
\epsfig{figure=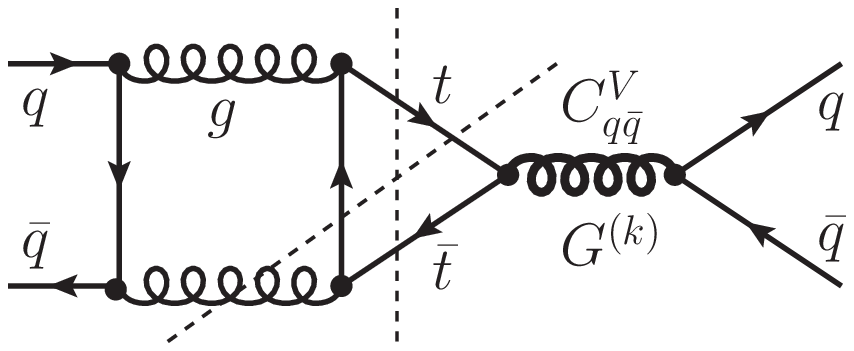,height=1in}
\end{center}
\caption{Charge-asymmetric contributions to $\sigma_a(p\bar p\rightarrow t\bar t)$ via the exchange of Kaluza-Klein gluons $G^{(k)}$ at leading order (left) and next-to-leading order (right).
\label{fig:nplo-npnlo}}
\end{figure}

\section{Top-quark pair production}
What are the resultant effects of KK gluons on the forward-backward asymmetry in top-quark pair production? The charge-asymmetric (-symmetric) cross section defined in Eq.~(\ref{eq:afbt}) is given by
\begin{equation}\label{eq:sigma}
\sigma_{a(s)} = \frac{\alpha_s}{m_t^2} \, \sum_{q} \int_{4m_t^2}^s
\frac{d \hat s}{s} \, \Big [ f\hspace{-0.4em}f_{q\bar q}\big( \hat s/s,\mu_f\big) - (+)  f\hspace{-0.4em}f_{\bar{q} q}\big( \hat s/s,\mu_f\big) \Big ] \, A_{q\bar q}\,(S_{q\bar q})\,,
\end{equation}
where $\hat s$ denotes the partonic CM energy, and the sum is taken over all quark flavors $q$ inside the proton. The parton luminosities $f\hspace{-0.4em}f_{q\bar q}\big( \hat s/s,\mu_f\big)$ as well as the hard-scattering kernels $A_{q\bar q}$ and $S_{q\bar q}$ in QCD are defined in \cite{Bauer:2010iq}. RS contributions at leading order (LO) arise from the interference of the $s$-channel exchange of a KK gluon with the tree-level gluon diagram. For the charge-asymmetric contribution, this is shown on the left-hand side of Figure~\ref{fig:nplo-npnlo}. The corresponding hard-scattering kernels for inclusive symmetric and asymmetric $t\bar t$ production are readily computed using the effective Lagrangian in Eq.~(\ref{eq:leff}),
\begin{equation}\label{eq:kernelLO}
S_{q\bar q}^{\rm LO} = \frac{\hat s}{M_{\rm{KK}}^2}\frac{\rho}{216}\sqrt{1-\rho}\left(2+\rho\right)\,\tilde{C}_{q\bar q}^V\,,\qquad A_{q\bar q}^{\rm LO} = \frac{\hat s}{M_{\rm{KK}}^2}\frac{\rho}{144}\left( 1-\rho \right)\,\tilde{C}_{q\bar q}^A\,,
\end{equation}
with $\rho = 4m_t^2/\hat{s}$ and $M_{\rm{KK}}$ in units of $1\,\rm{TeV}$. Referring to our considerations in Section~\ref{sec:RSfa}, we expect significant contributions of KK gluons to the cross section $\sigma_{t\bar t}$, which is sensitive to the vector coefficient $\tilde{C}_{q\bar q}^V=\mathcal{O}(1)$. Tree-level effects on the charge-asymmetric amplitude, however, are negligibly small due to the strongly suppressed axial-vector coefficient $\tilde{C}_{q\bar q}^A=\mathcal{O}(10^{-3})$. Thus there is no large forward-backward symmetry at tree level.

The suppression can be lifted by going to NLO, after paying the price of a loop factor $\alpha_s/(4\pi)$. The interference of a tree-level KK gluon exchange with a QCD box diagram, shown on the right-hand side of Figure~\ref{fig:nplo-npnlo}, has the same topology as the leading contribution to the asymmetry in QCD. The charge-asymmetric amplitude thus involves the unsuppressed vector coefficient $C_{q\bar q}^V$. The asymmetric hard-scattering kernel at NLO is given by
\begin{equation}\label{eq:kernelNLO}
A_{q\bar q}^{\rm NLO} = \frac{\alpha_s}{4\pi}\frac{\hat s}{M_{\rm{KK}}^2}\frac{\tilde{A}_{q\bar q}}{16\pi}\,\tilde{C}_{q\bar q}^V\,,
\end{equation}
with the SM coefficient $\tilde{A}_{q\bar q} \equiv A_{q\bar q}^{(1)}/\alpha_s$ defined in \cite{Bauer:2010iq}. Roughly, these NLO vector contributions are dominant if the condition $\alpha_s/(4\pi)\cdot (1+c_{t_L}+c_{t_R}) \gtrsim L \exp[L(1+c_{q_L}+c_{q_R})]$ is fulfilled by the bulk mass parameters. For parameter sets that reproduce the quark masses and mixings, the NLO contributions to $\sigma_a$ exceed the LO contributions by a factor of about $100$.

\section{Discussion and conclusions}
\begin{figure}[t]
\begin{center}
\epsfig{figure=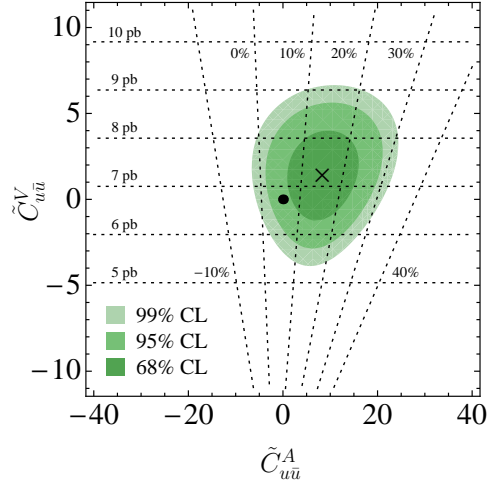,height=2.5in}
\end{center}
\caption{Constraints from combined fit to $\sigma_{t\bar t}$, $(d\sigma_{t\bar t}/dM_{t\bar t})^>$, and $A_{\rm{FB}}^{t,p\bar p}$ in the $\tilde{C}_{u\bar u}^A-\tilde{C}_{u\bar u}^V$ plane. The best fit point $(8.3,1.4)$ is marked by a cross, the SM $(0,0)$ by a black dot. The horizontal (vertical) lines correspond to the theoretical value of the total cross section $\sigma_{t\bar t}$ (of the forward-backward asymmetry $A_{\rm{FB}}^{t,p\bar p}$).
\label{fig:CACV}}
\end{figure}

After having assessed the size of vector and axial-vector RS contributions in top-quark pair production, we discuss the numerical effects on the observables. Using Eqs.~(\ref{eq:sigma}), (\ref{eq:kernelLO}), and (\ref{eq:kernelNLO}), and focussing on the dominant contributions from $u\bar u$ initial states, one obtains \footnote{For details on the inputs and the numerical calculation, please consult \cite{Bauer:2010iq}.}
\begin{eqnarray}\label{eq:ttbarobs}
 (\sigma_{t\bar t})_{\rm RS} & = & \left [ 1 + 0.053 \hspace{0.5mm} \tilde C_{u \bar u}^V \right ]
  \left ( 6.73^{+0.52}_{-0.80} \right ) {\rm pb} \,, \nonumber \\
  \left (\frac{d \sigma_{t\bar t}}{dM_{t\bar t}} \right )^{M_{t\bar t} \, \in \, [0.8, 1.4]\,\rm{TeV}}_{ \rm RS} & = & \left [ 1 + 0.33 \hspace{0.5mm} \tilde C_{u \bar
      u}^V \right ] \left ( 0.061^{+0.012}_{-0.006} \right ) \frac{\rm fb}{\rm GeV} \,,\\
(A_{\rm{FB}}^t)_{\rm{RS}}^{p\bar p} & = & \left [ \frac{1 + 0.22 \, \tilde C_{u \bar u}^A + 0.034 \,
    \tilde C_{u \bar u}^V}{1 + 0.053 \, \tilde C_{u \bar u}^V} \right
] \! \left ( 5.6^{+0.8}_{-1.0} \right ) \% \,. \nonumber
\end{eqnarray}
Discarding the axial-vector contributions, proportional to $\tilde{C}_{u\bar u}^A = \mathcal{O}(10^{-3})$, we inspect the interplay of NP effects mediated by the vector coefficient $\tilde{C}_{u\bar u}^V$. The corrections with respect to the SM prediction are at the percent level for all observables. Outstanding is the large correction to the cross section at high $M_{t\bar t}$ of about $30\%$ for $\tilde{C}_{u\bar u}^V=\mathcal{O}(1)$, compared to a moderate $5\%$ effect in the total cross section. This behavior is understood by noticing the increase of the NP contribution in the effective theory as $M_{t\bar t}^2/M_{\rm{KK}}^2$ at LO, cf. Eq.~(\ref{eq:kernelLO}). The high-$M_{t\bar t}$ bin of the distribution $d\sigma_{t\bar t}/dM_{t\bar t}$ in Eq.~(\ref{eq:ttbarobs}) constrains the vector coefficient to $\tilde{C}_{u\bar u}^V \in [-3.4,3.5]$ at $95\%$ CL. From the total cross section $\sigma_{t\bar t}$, one obtains an allowed range of $\tilde{C}_{u\bar u}^V \in [-1.5,7.4]$. These model-independent constraints limit the vector corrections to the charge-asymmetric cross section $\sigma_a$ to $[-6,+8]\%$ and $[-10,+3]\%$ of the SM prediction, respectively. Notice that the NLO effects in the numerator of $(A_{\rm{FB}}^t)_{\rm{RS}}^{p\bar p}$ are over-compensated by the charge-symmetric LO effects in the normalization. In the case of a positive vector coefficient $\tilde{C}_{u\bar u}^V$, the forward-backward asymmetry is thus always decreased with respect to its QCD value. The constraints on the effective couplings $C_{u\bar u}^V$ and $C_{u\bar u}^A$ are summarized in Figure~\ref{fig:CACV}, where we show a combined fit to the $t\bar t$ observables $\sigma_{t\bar t}$, $(d\sigma_{t\bar t}/dM_{t\bar t})^>$, and $A_{\rm{FB}}^{t,p\bar p}$.
It is clearly visible that the forward-backward asymmetry (vertical dashed lines) cannot be increased beyond $A_{\rm{FB}}^{t,p\bar p} = 5.8\%$ ($6.0\%$) at the $95\%$ (99\%) CL by vector contributions alone. To reach the best-fit point $(\tilde{C}_{u\bar u}^A,\tilde{C}_{u\bar u}^V)=(8.3,1.4)$, large axial-vector contributions at tree level are required. \footnote{To satisfy the measured asymmetry at high $M_{t\bar t}$, axial-vector contributions have to be even larger.} In the RS model with flavor anarchy, one typically has $\tilde{C}_{q\bar q}^{V}\approx +0.5$ in the regime of perturbative Yukawa couplings. The resulting absolute correction to the asymmetry amounts to $\delta (A_{\rm{FB}}^t)_{\rm{RS}}^{p\bar p} = -0.05\%$, assuming a KK scale of $M_{\rm{KK}}=1\,\rm{TeV}$. For this scale, the first KK excitation of the gluon exhibits a mass of around $2.5\,\rm{TeV}$. Notice that for new particles with masses below about $2\,\rm{TeV}$, the treatment in terms of an effective theory does not apply any longer and width effects have to be taken into account.

In summary, top-quark pair production in the RS model with flavor anarchy is essentially forward-backward symmetric: axial-vector contributions from KK gluons at tree level are strongly suppressed, because the light quarks are localized in the UV regime of the extra dimension. At NLO, a charge asymmetry arises from vector contributions, but these effects are over-compensated by the simultaneous LO contributions in the symmetric cross section, which normalizes the forward-backward asymmetry. On general grounds, it is therefore not possible to achieve a large forward-backward asymmetry from vector contributions alone. The asymmetry has to be generated at tree level either by NP with large axial-vector couplings in the $s$ channel or flavor-changing couplings in the $t$ channel. In the model at hand, the forward-backward asymmetry is slightly decreased with respect to the SM value. A larger asymmetry in the RS framework can theoretically be obtained by moving the localization of the light quarks more towards the IR regime. The prize to pay is to give up the appealing feature of explaining the quark mass hierarchy exclusively by localization in the extra dimension. The LHC starts to probe charge-asymmetric top-quark pair production. However, KK gluons - if existing - will rather show up as resonances in the symmetric $t\bar t$ cross section. They naturally generate a large top-quark forward-backward \emph{symmetry}.

\section*{Acknowledgments}
It is a pleasure to thank the organizers of \emph{Les Rencontres de Moriond 2011 EW} for an interesting and inspiring conference, as well as Martin Bauer, Florian Goertz, Uli Haisch, and Torsten Pfoh for an enjoyable collaboration on this subject. Many thanks to Paul Archer and Uli Haisch for proofreading the manuscript. This research is supported by the Helmholtz-Institut Mainz.

\end{document}